\documentclass[showpacs, twocolumn, preprintnumbers,amsmath,superscriptaddress, amssymb, prl]{revtex4-1}
\usepackage{graphicx}
\usepackage{amsmath, amsthm, amssymb}
\usepackage{latexsym}
\usepackage{amssymb}
\usepackage{bm}
\usepackage{dcolumn}
\usepackage{epstopdf}
\usepackage[usenames, dvipsnames]{xcolor}
\usepackage{braket}
\usepackage[normalem]{ulem}

\newcommand{\UP}{UPt$_3$}
\newcommand{\lam}{$\lambda$}

\begin{document}
\title{Nodal Gap Structure and Order Parameter Symmetry of the Unconventional Superconductor \UP\ }
\date{\today}

\author{W. J. Gannon}
\altaffiliation[Current address: ]{Department of Physics and Astronomy, Stony Brook University, Stony Brook, NY USA}
\author{W. P. Halperin}
\email{w-halperin@northwestern.edu}
\affiliation{Department of Physics and Astronomy, Northwestern University, Evanston, IL 60208 USA}

\author{C. Rastovski} 
\affiliation{Department of Physics, University of Notre Dame, Notre Dame, IN 46556 USA}

\author{K. J. Schlesinger}
\altaffiliation[Current address: ]{Department of Physics, University of California, Santa Barbara, CA 93106 USA}
\noindent

\affiliation{Department of Physics, University of Notre Dame, Notre Dame, IN 46556 USA}

\author{J. Hlevyack}
\affiliation{Department of Physics, Loyola University Chicago, Chicago, IL 60626 USA }
\altaffiliation[Current address: ]{Department of Physics, University of Illinois Urbana-Champaign, Urbana, IL 61801 USA}
\author{M. R. Eskildsen} 
\affiliation{Department of Physics, University of Notre Dame, Notre Dame, IN 46556 USA}
\noindent
\author{A.B. Vorontsov}
\affiliation{Department of Physics, Montana State University, Bozeman, MT 59717 USA}
\author{J. Gavilano}
\author{U. Gasser}
\author{G. Nagy}
\affiliation{Laboratory for Neutron Scattering, Paul Scherrer Institut,  CH-5232, Villigen, Switzerland}

\begin{abstract}
Spanning a broad range of physical systems, complex symmetry breaking is widely recognized as  a hallmark of competing interactions. This is exemplified in superfluid $^3$He which has multiple thermodynamic phases with spin and orbital quantum numbers $S =1$ and $L =1$, that emerge on cooling  from a nearly ferromagnetic Fermi liquid.  The heavy fermion compound \UP\ exhibits similar behavior clearly manifest in its multiple superconducting phases.  However, consensus as to its order parameter symmetry has remained elusive.  Our small angle neutron scattering measurements indicate a  linear temperature dependence of the London  penetration depth characteristic of nodal structure of the order parameter.  Our theoretical analysis is consistent with  assignment of its symmetry to an $L = 3$ odd parity state for which one of the three thermodynamic phases in non-zero magnetic field is chiral.
\end{abstract}
\maketitle

\noindent
{\bf 1. Introduction}\\

\noindent
Recent interest in topological superconductors has focused attention on materials that exhibit chiral symmetry, or have been proposed to exhibit chiral symmetry, including Sr$_2$RuO$_4$, $^3$He, and \UP~\cite{Sauls_NJP_2009}.  In particular, the heavy fermion  compound \UP ~\cite{Stewart_PRL_1984, Joynt_RevModPhys_2002} has attracted  theoretical attention~\cite{Graf_PRB_2000, Norman_chapter_2013, Tsutsumi_JPSJ_2012} in part as a consequence of conflicting experimental reports on the nature of its unconventional superconducting state.  For example, the observation of Pauli limiting in the upper critical field~\cite{Shivaram_PRL_1986} appears to be incompatible  with temperature independence of the Knight shift~\cite{Tou_PRL_1996, Tou_PRL_1998}.  Josephson tunneling interference measurements~\cite{Strand_PRL_2009} and measurements of the polar Kerr effect~\cite{Schemm_Science_2014} provide evidence for an order parameter that is chiral in the B-phase.  However, recent directional thermal conductivity experiments are interpreted otherwise~\cite{Tsutsumi_JPSJ_2012}.  Here we use small angle neutron scattering (SANS) from the vortex lattice (VL) to provide  a bulk probe of the temperature dependence of the penetration depth, obtaining evidence for the nodal structure of the order parameter in the B-phase supporting its identification as an odd parity, chiral state, with E$_{2u}$ symmetry consistent with theory~\cite{Sauls_AdvPhys_1994}. 

\begin{figure}
\includegraphics[width=80mm]{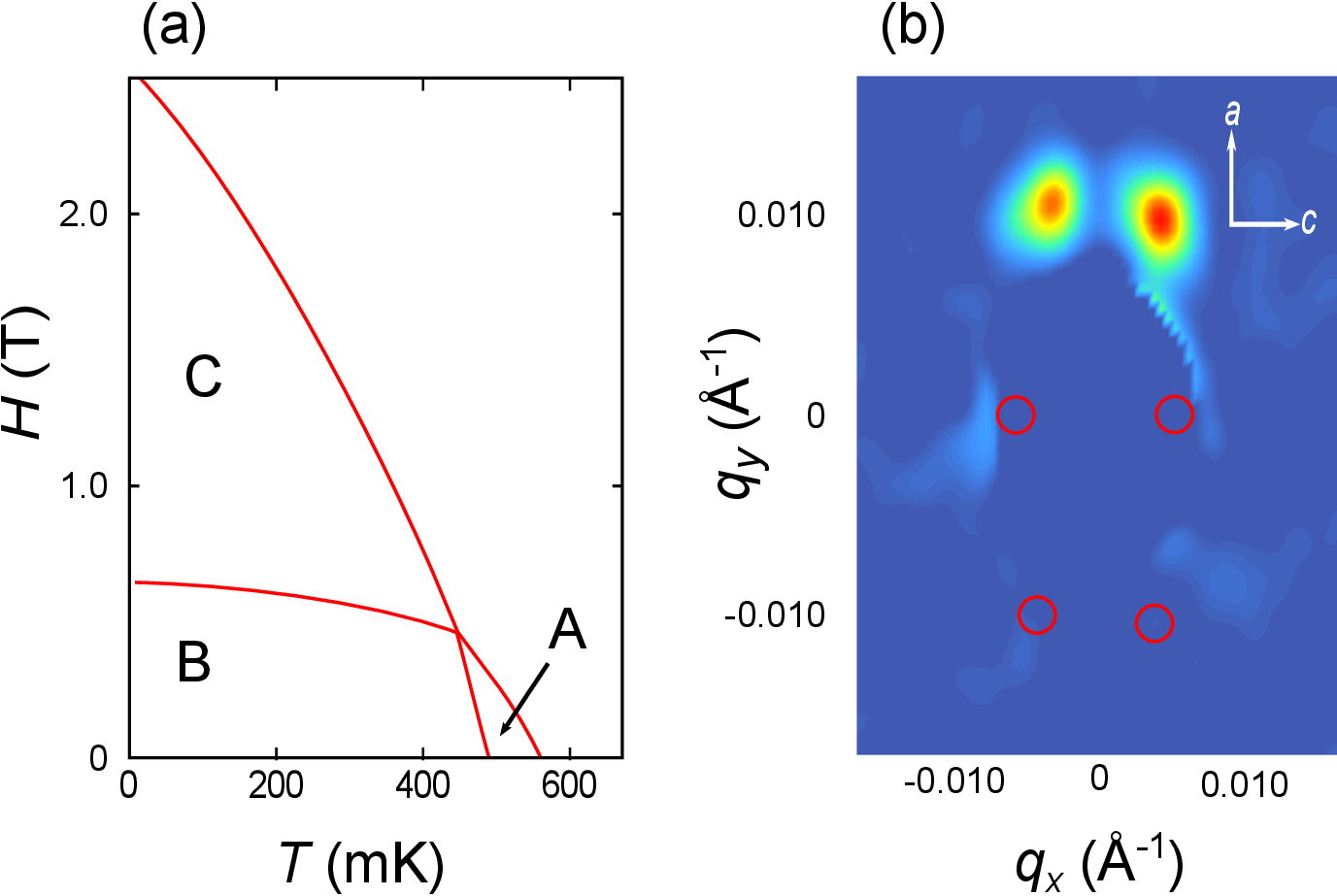}
\caption{\label{Fig1}  Phase diagram of \UP\ and its vortex diffraction pattern. ({\bf a}) A schematic of the phase diagram for the three vortex phases, A, B, and C, of \UP\ for $H||a^*$.  The normal to superconducting transition is $T_c$.  The  transitions between phases are $T_{AB}$ and $T_{BC}$.  ({\bf b}) An example of a diffraction pattern, measured at $\approx$ 50 mK for $H=0.3$ T with $H||a^*$. The Bragg condition was only satisfied for diffraction peaks above the beam center.  By symmetry, there are two peaks below the beam center and 2 peaks on the horizontal, indicated by red circles.  The \UP\ crystal axes are shown as white arrows.}
\end{figure}

One of the most striking properties of \UP\ is the fact that the $H-T$ superconducting phase diagram has three distinct superconducting vortex phases shown in Fig.~\ref{Fig1}a, conventionally labeled A, B, and C.  Experiments and theory demonstrate that this phase diagram can only be explained by an unconventional superconducting order parameter~\cite{Joynt_RevModPhys_2002}, a close parallel to superfluid $^3$He.  However, a complete theoretical description of the superconducting state of \UP\ has not been settled, and there are several candidate models that can account for the material's unusual physical properties.  The order parameter structure that is consistent with a number of experiments is an odd-parity, $f$-wave (L=3) orbital state of E$_{2u}$ symmetry~\cite{Sauls_AdvPhys_1994, Graf_PRB_2000}.  However, with some success, comparisons with experiment have also been made for an even-parity,  $d$-wave (L=2) orbital state of E$_{1g}$ symmetry~\cite{Park_PRB_1996}.  Both of these order parameters are chiral and break time reversal symmetry in the low temperature B-phase in contrast to a recent proposal~\cite{Tsutsumi_JPSJ_2012} for an odd-parity, $f$-wave (L=3) model  with E$_{1u}$ orbital symmetry which  is non-chiral and time reversal symmetric in the B-phase.

All of these order parameters have nodes in the superconducting energy gap, each with different nodal structure in the three vortex phases.  Consequently, it is of particular importance to explore physical properties that are directly linked to this nodal structure and that are sensitive to the gap dispersion at the nodes.  Using SANS from the VL, we have measured the temperature dependence of the components of the London penetration depth, \lam$_i$(\textit{T}), that probe the gap nodal structure along the principal directions of the crystal~\cite{Prozorov_SST_2006, Furukawa_PRB_2011} finding  linear behavior in the low temperature limit.  
%abv 1 
%Our quasi-classical calculations using a 
Our calculations using the quasiclassical  
Green's function approach and an ellipsoidal Fermi surface are consistent  with the \lam$_i$(\textit{T}) data over a wide range of temperature for an order parameter with E$_{2u}$ symmetry.  We compare our results for \lam$_i$(\textit{T}) with those from other experimental methods including ac-susceptibility~\cite{GrossAlltag_ZPB_1991, Signore_PRB_1995} and muon spin rotation ($\mu$SR)~\cite{Broholm_PRL_1990, Yaouanc_JPCM_1998}. \\

%*********************************************************************
%***********************************   FIGURE 1   *******************%*********************************************************************

%*********************************************************************
%***********************************   FIGURE 2   *******************%*********************************************************************
\begin{figure}
\includegraphics[width=80mm]{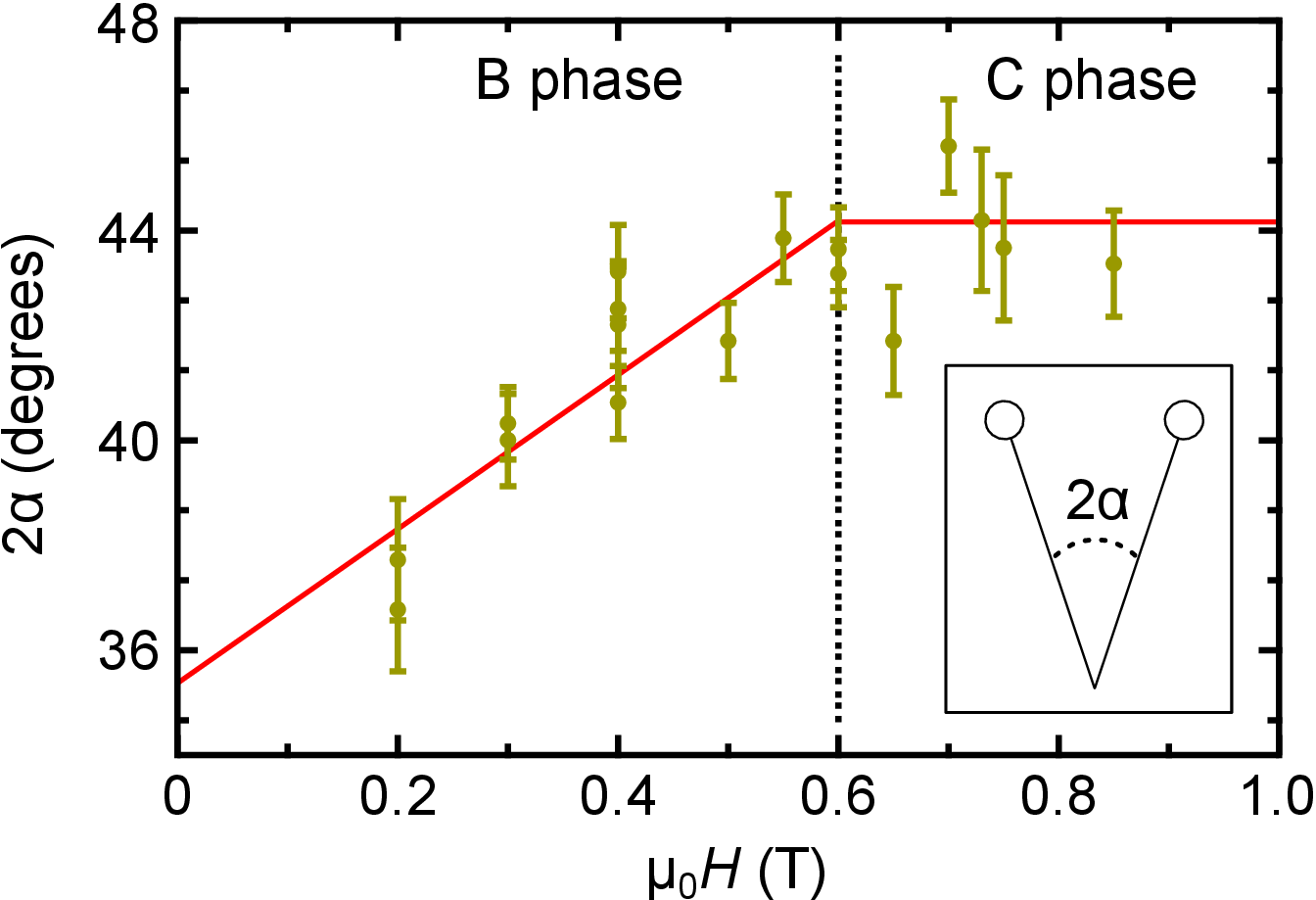}
\caption{\label{Fig2} The VL opening angle in B and C-phases. The opening angle  $2\alpha$ is defined in the inset, shown as a function of applied magnetic field along the $a^*$-axis measured at $\approx$ 50 mK. The B to C-phase transition is given by the vertical dashed line at $H=0.6$ T.  Red lines are guides to the eye.}
\end{figure}

\noindent
{ \bf 2. Experimental Methods}\\

\noindent
Our sample consists of  a high-quality, 15 g single crystal (RRR $> 600$), cut into two pieces, and is described by Gannon \textit{et al.}~\cite{Gannon_PRB_2012}.  The \UP\ crystals were  co-aligned, fixed with silver epoxy to a copper cold finger,  and mounted to the mixing chamber of a dilution refrigerator with the crystal $a$-axis vertical and the $c$  and $a^{*}$-axes in the horizontal plane.  Rotation of the dilution insert allowed easy reorientation of the $a^*$ or $c$-axes to be parallel to the magnetic field and neutron beam inside a horizontal superconducting magnet on the SANS-I and SANS-II beamlines at the Paul Scherrer Institut in Villigen, Switzerland.  For measurements on SANS-I, the neutron wavelength was 6 \AA\ with 11 m of collimation and detector to sample distance between 16 and 20 m.  For measurements on SANS-II, 9 \AA\ neutrons were used with 6 m of collimation and the detector to sample distance was 6 m.\\

\noindent
{\bf 3. Results}\\

\begin{figure}
\includegraphics[width=80mm]{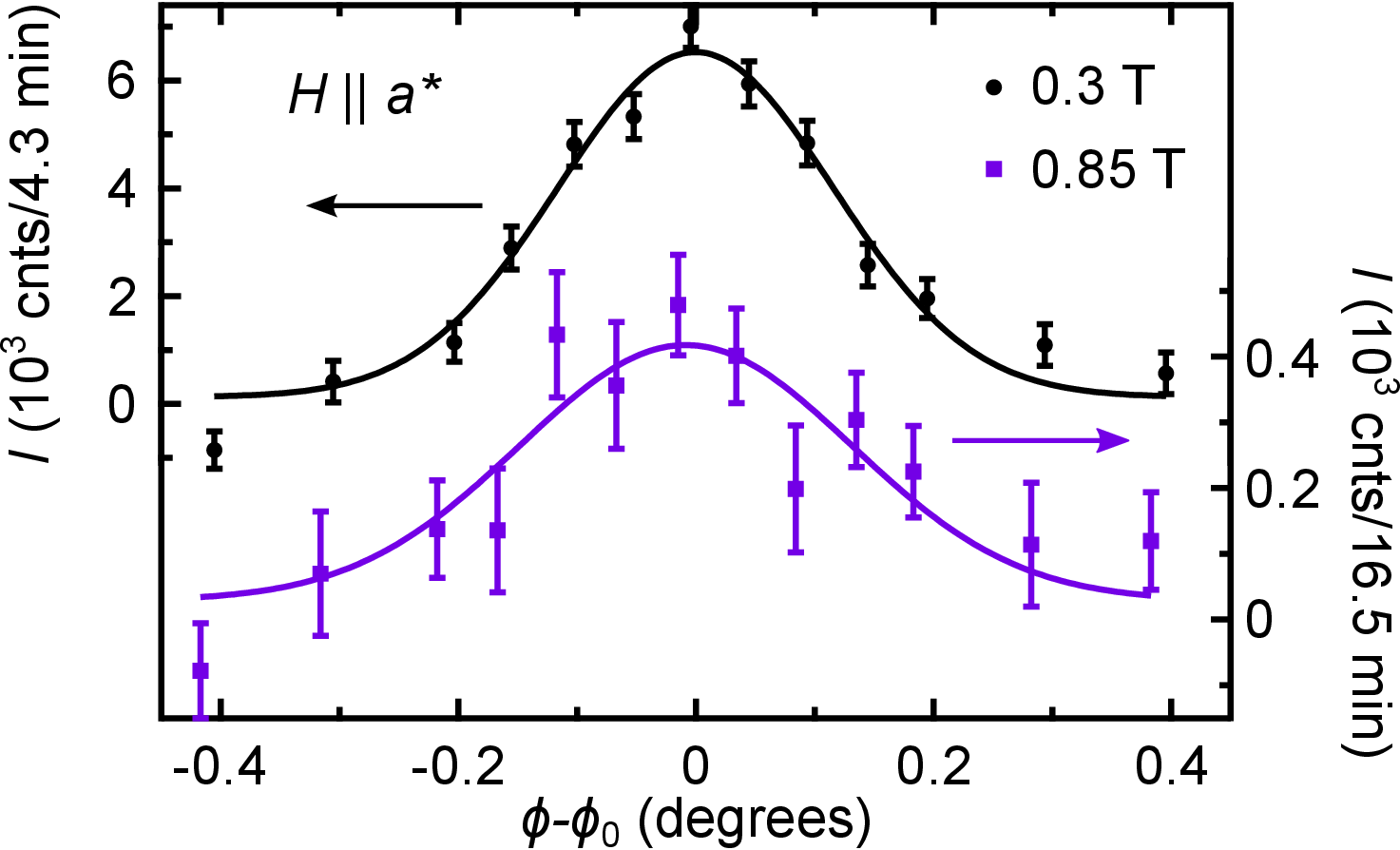}
\caption{\label{Fig3}  Rocking curves.  Examples of rocking curves are presented at $\approx$ 50 mK for $H=0.3$ T (black circles) and 0.85 T (purple squares) with fields along the $a^*$-axis.   Fits to each rocking curve are gaussian and the $\phi$ axis is shifted to the curve center, $\phi_0$. }
\end{figure}

\noindent
A typical result of a SANS diffraction pattern from \UP, is shown in Fig.~~\ref{Fig1}b with magnetic field $H=0.3$ T parallel to the $a^*$-axis.  In the present work we have made measurements of many similar patterns as a function of temperature and magnetic field, Fig.~\ref{Fig2}.   Only the first order Bragg reflections were observed as in Fig.~\ref{Fig1}b, since \UP\ has relatively long penetration depths. The diffraction patterns were constructed from a superposition of scattering images measured at different rocking angles $\phi$ about the horizontal, Fig.~\ref{Fig3}, where the Bragg condition was satisfied above the beam center.  For all data discussed here background scattering measured in zero applied field was subtracted.  The diffraction pattern shown in Fig.~\ref{Fig1}b is that of a distorted hexagonal vortex lattice (VL), similar to, but more anisotropic than, previous SANS measurements in this orientation~\cite{Kleiman_PRL_1992, Yaron_PRL_1997}.  By symmetry, there are four additional peaks indicated by red circles that were not imaged in the hexagonal domain in Fig.~\ref{Fig1}b, but they were directly observed at higher fields.  The symmetry of the diffraction pattern is the same as that of the real space VL, rotated by 90 degrees with a rescaling of the axes.  The distortion of the VL from a perfect hexagon is a result of penetration depth anisotropy in the plane perpendicular to $a^*.$  When a 0.2 T field is applied parallel to the $c$-axis, a perfect hexagonal VL is seen within our resolution, in agreement with previous measurements for that orientation at a similar field~\cite{Huxley_Nature_2000}.

Fig.~\ref{Fig2} shows the opening angle 2$\alpha$ of the VL, defined in the inset, as a function of applied magnetic field for $H||a^*$.  For the data shown here the VL was prepared by reducing the magnetic field at constant temperature $\approx 50$ mK from above the upper critical field, $H_{c2}$.  Then, a damped oscillation with initial magnitude 0.02 T was performed around the final measurement field.  The motivation for preparing the VL with this field history was to produce the superconducting state with an equilibrium order parameter orientation and to ensure that the VL was in the ground state~\cite{Huxley_Nature_2000, Das_PRL_2012}.

Our $2\alpha$ data can be best described as having a  linear field dependence in the B-phase that becomes field independent in the C-phase, where the B to C-transition occurs between $H= 0.5~\mathrm{and}~0.6$ T in this field orientation, inferred from the phase diagram of Adenwalla \textit{et al.}~\cite{Adenwalla_PRL_1990}.  The field dependence to the opening angle indicates that non-local corrections to the London theory are signifyicant~\cite{Sauls_AdvPhys_1994, White_PRB_2011, Sauls_comm_2013}.  A change in the field dependence of the opening angle at the B to C-transition was also reported by Yaron \textit{et al.}~\cite{Yaron_PRL_1997}.

The intensity in a diffraction peak is related to the Fourier transform of the local field variations from the real space VL.  Systematic measurements as a function of rocking angle were made to produce rocking curves such as displayed in Fig.~\ref{Fig3}.   We measured the first order Fourier component of the diffraction, called the form factor $|h_{1}|$, expressed as,
 \begin{equation}\label{IntegratedIntensity}
|h_{1}|^{2} = R\frac{16\Phi_{0}^{2}q}{2\pi\gamma^{2}\lambda^{2}_{n}t} .
\end{equation}

%*********************************************************************
%***********************************   FIGURE 3   *******************%*********************************************************************

The form factor is calculated from the reflectivity, $R$, equal to the integrated intensity of a rocking curve  multiplied by cos\,$\alpha$ (the Lorentz factor), divided by the incident neutron flux.  In Eq.~\ref{IntegratedIntensity}, $\Phi_{0}=2.07$x$10^5\,$T$\,\cdot$\AA$^2$ is the magnetic flux quantum; $q$ is the magnitude of the scattering vector of the reflection being measured; the gyromagnetic ratio of the neutron is $\gamma=1.91$; $\lambda_n$ is the incident neutron wavelength; and $t$ is the effective sample thickness which we have taken to be 3.9 mm -- the equivalent thickness of a uniform sample with the same width, height, and volume as our sample.  The field dependence of our measurements of rocking curve widths do not show the sudden broadening at the B-C phase transition reported by Yaron \textit{et al.}~\cite{Yaron_PRL_1997}.  All of our rocking curves are only $\sim20$\% broader than the resolution limit for our experiments.  We also do not see a change in slope of the field dependence of $|h_1|$ at the B-C transition as reported earlier~\cite{Yaron_PRL_1997}.  It is likely that absence of these effects can be attributed to the higher quality of our crystal and the oscillatory field procedure which we have used to overcome flux pinning.

%*********************************************************************
%***********************************   FIGURE 4  *******************%*********************************************************************
\begin{figure}
\includegraphics[width=80mm]{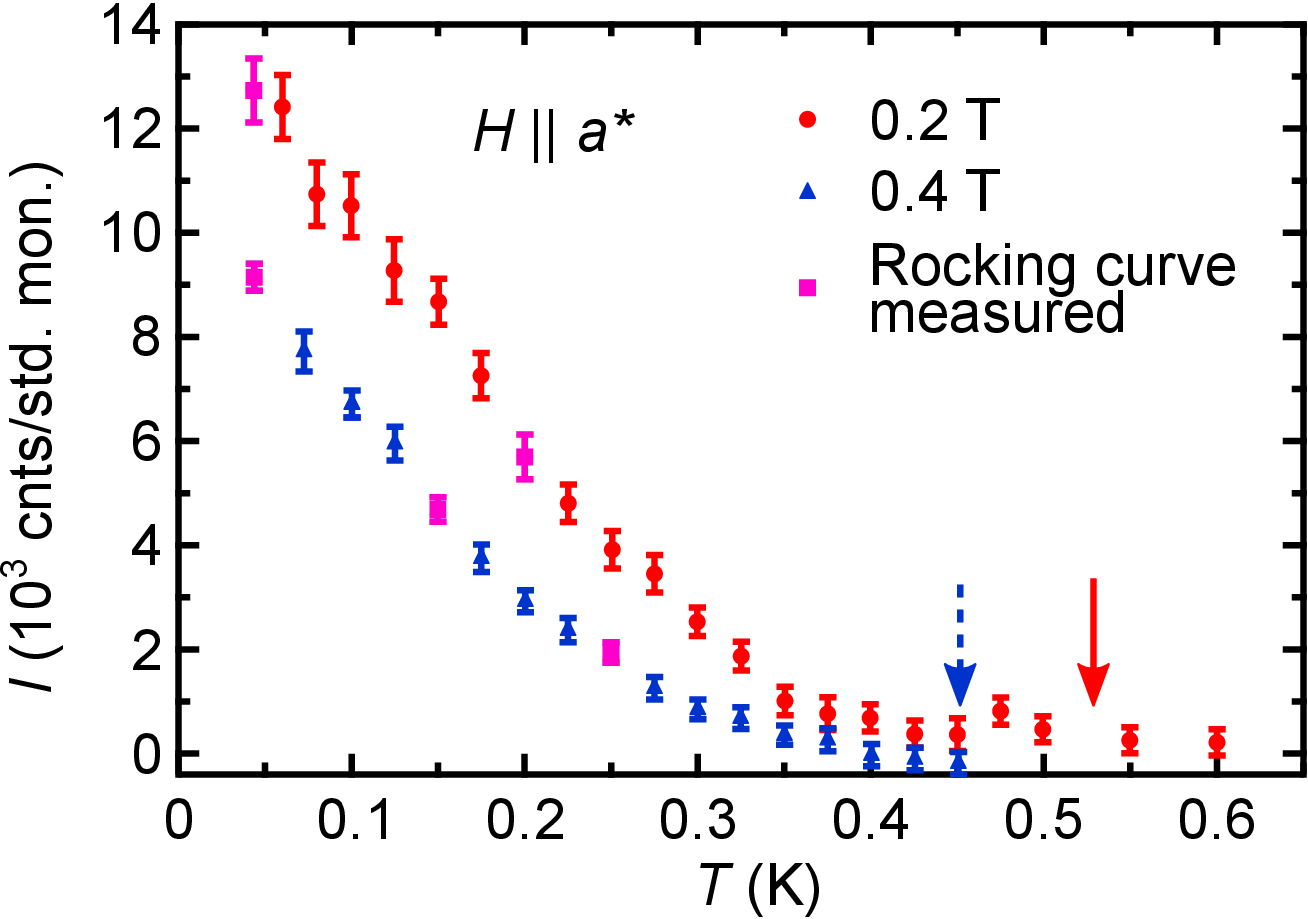}
\caption{\label{Fig4}  Scattered neutron intensity.  The rocked-on SANS intensity as a function of applied magnetic field is shown for fields along the $a^*$-axis at $H=0.2$ T (red circles) and 0.4 T.  The solid red arrow indicates $T_c = 520$ mK at 0.2 T.  The dashed blue arrow indicates $T_c = 450$ mK at 0.4 T. }

\end{figure}

%*********************************************************************
%***********************************   FIGURE 5  *******************%*********************************************************************
\begin{figure}
\includegraphics[width=80mm]{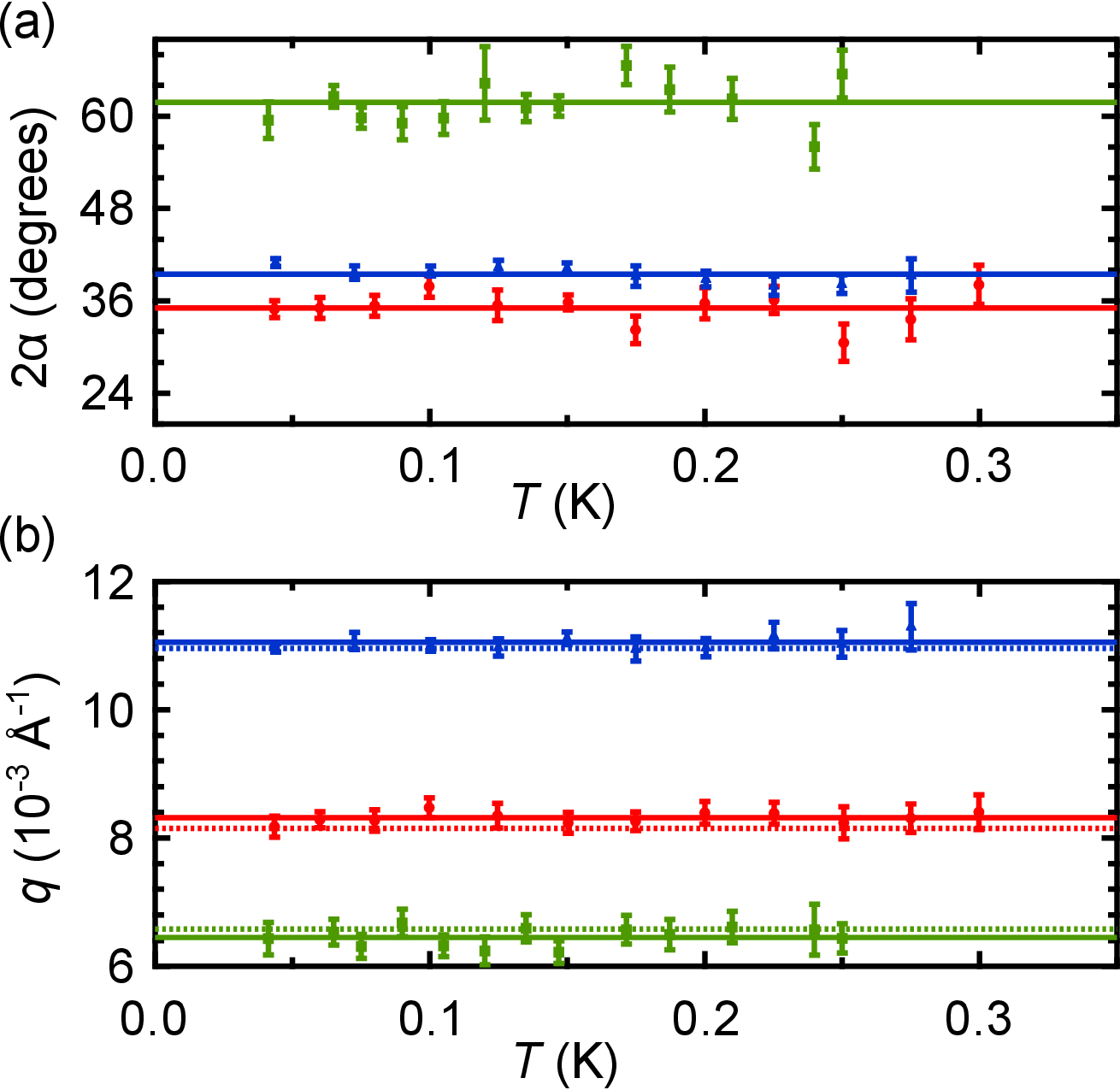}
\caption{\label{Fig5} Temperature dependence of the opening angle  and scattering vector. ({\bf a}) The opening angle $2\alpha(T)$ and ({\bf b}) the magnitude of the scattering wave vector  $q(T)$ is shown as a function of temperature.  For each panel, the data is given for fields along the $a^*$-axis at 0.2 T (red circles) and 0.4 T (blue triangles). For $H||c$ at 0.2 T  the data (green squares)  are within error bars of being a perfectly hexagonal VL.  The solid lines show the average for each data set.  The dashed lines in panel ({\bf b}) show $q$ calculated using the average values of 2$\alpha$ taken from panel (a), assuming  single flux quantization and $B=\mu_oH$.}
\end{figure}

In the London theory the form factor is related to the material properties through the magnetic penetration depth \lam.  The form factor for an isotropic superconductor is given by, 
\begin{equation}\label{London1}
|h_1|=\frac{B}{1+\lambda^{2}q^{2}}e^{-c\xi^{2}q^{2}}
\end{equation}
where $\xi$ is the superconducting coherence length and c is a constant, typically taken to be $\frac{1}{2}$.   The fractional part of  Eq.~\ref{London1} comes directly from  the London equations~\cite{Kogan_PLA_1981}.  The exponential factor is a correction to the London theory to account for the non-zero extent of the vortex cores~\cite{Yaouanc_PRB_1997, Densmore_PRB_2009}.  This simple gaussian model for the core correction with the constant $c=\frac{1}{2}$ has been found to be more accurate than  more sophisticated models~\cite{Densmore_PRB_2009}.  Nonetheless, the temperature dependence of the penetration depth is not sensitive to the choice of this correction and its exact value is immaterial to the conclusions in the present work.  For an anisotropic superconductor  the form factor can be expressed in terms of the principal values of the penetration depth \lam$_i$ corresponding to currents flowing along each of the principal directions of the crystal,  with $i= 3$ for currents along the $c$-axis.  Measuring the form factor therefore provides a direction-specific probe of the low lying excitations in the superconducting state sensitive to gap nodes~\cite{Prozorov_SST_2006}.

For uniaxial anisotropy, as for  \UP, $\lambda_1 = \lambda_2 \neq \lambda_3$ and the form factor for fields along the $a$ or $a^*$-axis  becomes~\cite{Kogan_PLA_1981},

\begin{equation}\label{London2}
|h_1|=\frac{B}{1+\lambda_1^{2}q^{2}\mathrm{sin}^{2}\alpha+\lambda_3^{2}q^{2}\mathrm{cos}^{2}\alpha}e^{-c\xi^{2}q^{2}},
\end{equation}

\noindent
where the \lam$_i$ are related to the corresponding diagonal components of the quasi-particle mass tensor m$_i$, and the opening angle $2\alpha$, through the relation,

\begin{equation}\label{London3}
\mathrm{tan}^2 \alpha\,\,=\,\,(m_3/3m_1)\,\,=\,\,(\lambda_3^2/3\lambda_1^2).
\end{equation}

If there is no variation in the VL geometry as a function of temperature, as demonstrated in Fig.~\ref{Fig5}, then the temperature dependence of the form factor given by Eq.~\ref{London2} reflects the temperature dependence of  \lam$_3$ where the denominator of Eq.~\ref{London2} simplifies to $1 + \tfrac{4}{3} \lambda_3^2 q^2 \cos^2 \alpha$.

We have made measurements of the temperature dependence of the VL scattering for magnetic fields along both the crystal $c$ and $a^*$-axes with the field reduced from above $H_{c2}$, followed by damped field oscillations before measurement at each temperature.  Rocking curves were obtained for each orientation at base temperature and at intermediate temperatures to determine that there was no broadening as temperature was varied.   The magnet and sample were rotated to the center of the rocking curve and the scattered intensity $I(T)$ was measured ``rocked-on'' ($\phi=\phi_0$) as a function of temperature, Fig.~\ref{Fig4}.

\begin{figure}
\includegraphics[width=70mm]{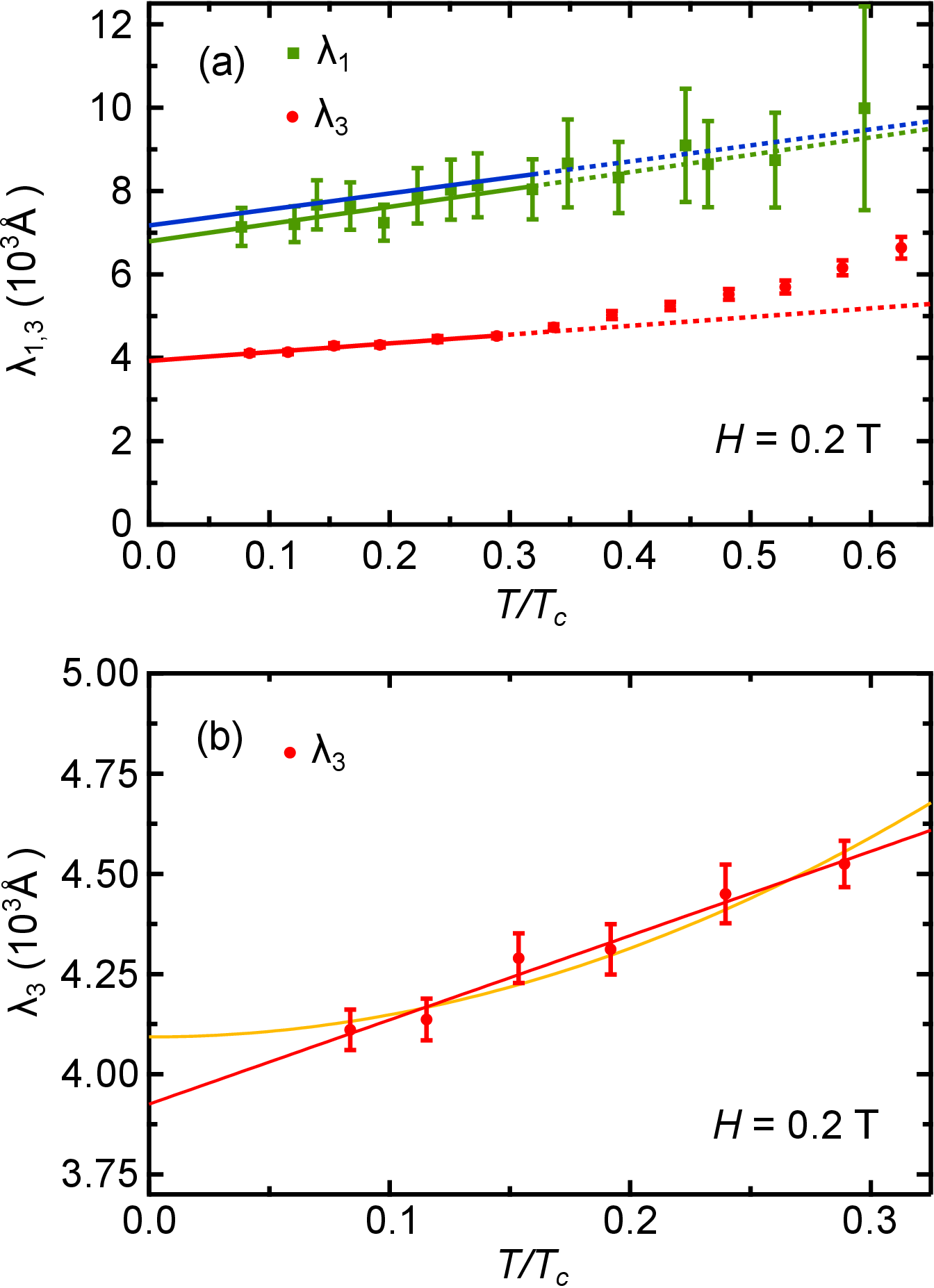}
\caption{\label{Fig6} The temperature dependence of the penetration depth.  ({\bf a}) $\lambda_1$ calculated with Eq.~\ref{London1} (green squares) and $\lambda_3$ (red circles, from Eq.~\ref{London2} ) over the whole temperature range with linear fits to each (green and red lines) from base temperature to $T/T_c=0.3$ in a magnetic field $H=0.2$ T.  The blue line is derived from  analysis of \lam$_1(T)$ together with the opening angle $\alpha(T)$ in the context of the London theory, Eq.~\ref{London3}.  The statistical accuracy of the measurements for \lam$_3$ is approximately the size of the data points. ({\bf b}) $\lambda_3$ 
in the low temperature region with linear (red) and quadratic (yellow) power law fits.  The linear fit is significantly better with $\chi^2$ favoring linear temperature dependence by a factor of 2.23 as compare with a quadratic fit. Additionally,  the temperature dependence over a broad range is consistent with our theoretical analysis, Fig.~\ref{Fig7} for quadratically dispersed point nodes along the $c$-axis that leads to a linear temperature dependence at low temperatures.}
\end{figure}

The opening angle and the scattering vector, taken directly from the diffraction pattern, are both temperature independent as shown in Fig.~\ref{Fig5}a,b.   The penetration depth anisotropy at low temperatures obtained directly from the opening angle is,  $\lambda_1/\lambda_3=1.83 \pm 0.04$ at $H=0.2$ T, giving a quasiparticle mass anisotropy of $m_1/m_3=3.34 \pm 0.13$. Using the average values of $\alpha$ and $q$, we calculated $|h_1|$ from Eq.~\ref{IntegratedIntensity} at $H=0.2$ T for each temperature and field orientation.    We determined \lam$_3(T)$, shown in Fig.~\ref{Fig6} from the simplified version of Eq.~\ref{London2} using our $|h_1|$ values for $H||a^*$, the average values of $\alpha$ and $q$ for this field orientation, and $\xi$=110 \AA~\cite{Shivaram_PRL_1986}.  Since the penetration depth is isotropic in the plane perpendicular to the $c$-axis, we used Eq.~\ref{London1} and our results for $|h_1|$ with $H||c$ to find \lam$_1(T)$.  \\

\noindent
{\bf 4. Penetration Depth at Low Temperature}\\

\noindent
The nodal structure of the order parameter is  evident from the VL scattering cross-section in its low temperature limiting behavior where the quasiparticle thermal excitation energies are much less than $k_BT_c$.  From earlier work, notably thermal conductivity and sound attenuation~\cite{Joynt_RevModPhys_2002} together with the theory~\cite{Sauls_AdvPhys_1994, Graf_PRB_2000}, this limiting low temperature region is $T/T_c  \lesssim 0.4$ which we conservatively take to be $T/T_c  \lesssim 0.3$.  We have compared linear and quadratic fits to the temperature dependence of $\lambda_3$ over this temperature range shown in detail in Fig.~\ref{Fig6}b.  Our data is consistent with a linear temperature dependence which provides a  significantly better description than quadratic behavior  as indicated by our chi-squared analyses for the fits shown in this figure.

%*********************************************************************
%***********************************   FIGURE 6   *******************%*********************************************************************

%*********************************************************************
%***********************************   FIGURE 7   *******************%*********************************************************************
\begin{figure}
\includegraphics[width=75mm]{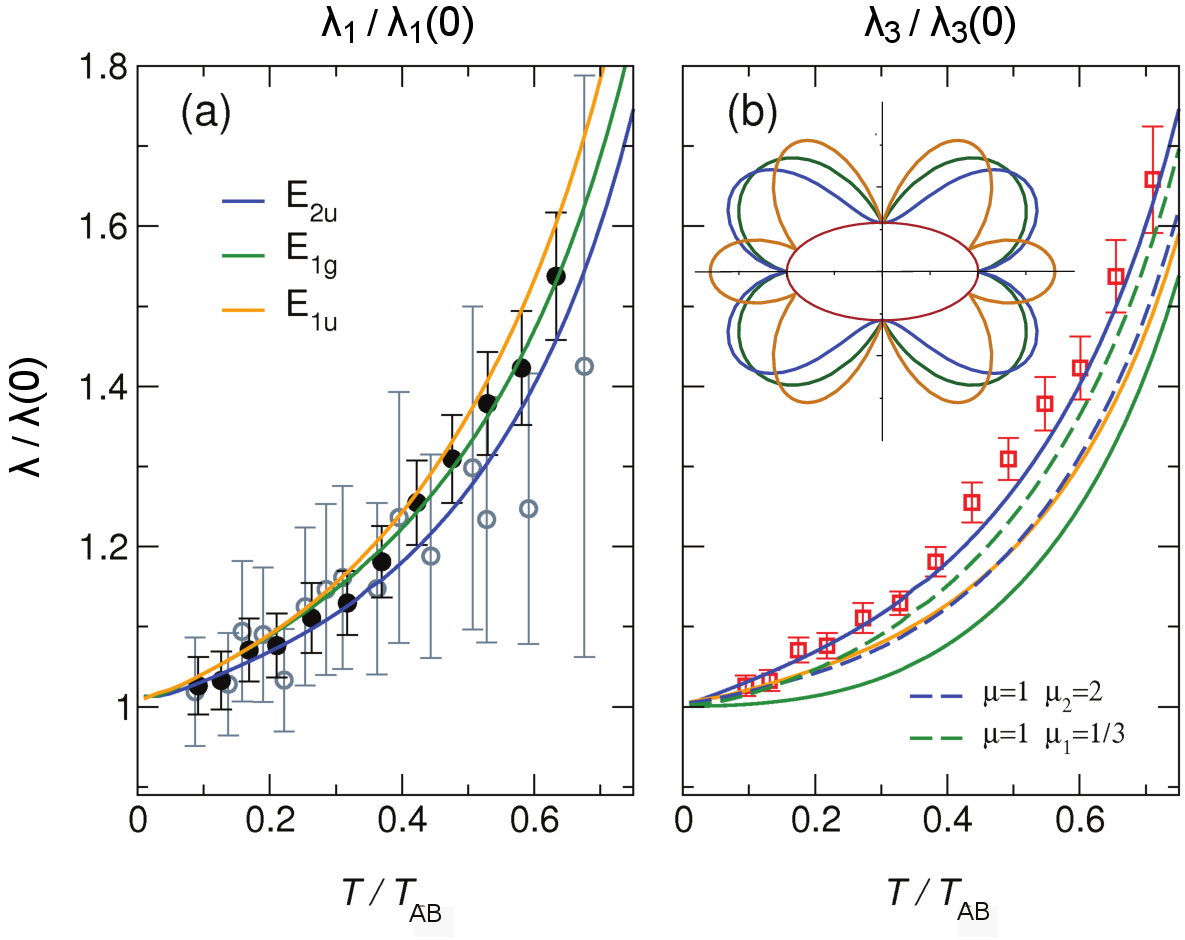}
\caption{\label{Fig7} Theoretical calculation of the penetration depth.  Comparison is made with three models for the symmetry of the order parameter E$_{2u}$, E$_{1g}$, and E$_{1u}$  ({\bf a}) The calculations of $\lambda_1(T)/\lambda_1(0)$ as a function of temperature are shown for each model order parameter symmetry and an ellipsoidal Fermi surface, (b) inset.  The open circles in ({\bf a}) are data from Fig.~\ref{Fig6}a labeled $\lambda_1$.  An independent and more accurate data set for $\lambda_1$ (solid circles) was obtained from the data for $\lambda_3$ combined with measurements of the opening angle as described in the text, Eq.~\ref{London3}. ({\bf b}) The calculations of $\lambda_3(T)/\lambda_3(0)$ give the temperature dependence of the penetration depth for currents along the $c$-axis.  The linear behavior at low temperature for the E$_{2u}$ state is a consequence of the quadratic dispersion of the energy gap for the polar nodes. Solid curves are for E$_{2u}$ (blue), E$_{1g}$ (green), and E$_{1u}$ (yellow). Dashed lines show results for different nodal openings $\mu_1$ (E$_{1g}$) and $\mu_2$ (E$_{2u}$).  ({\bf b} Inset) Gap profiles for the three candidate order parameters in the B-phase on an ellipsoidal Fermi surface.}
\end{figure}

For $\lambda_1(T)$, $i.e.\,H||c$,   the accuracy of the data is less than for $\lambda_3(T)$ since the corresponding penetration depth is larger  and the spatial variations of the local magnetic field from which the neutrons are scattered are much smaller.  However, we have independent information from the diffraction pattern resident in our measurement of the opening angle $\alpha(T)$.  Within the context of the London theory we can determine $\lambda_1(T)$ from Eq.~\ref{London3}.
We plot  this determination of  $\lambda_1(T)$ in Fig.~\ref{Fig6}a as a blue line, which is also linear in temperature just as is $\lambda_3(T)$.  Our extrapolations to zero temperature with linear fits to the data give:  
$\lambda_1(0) = 6,800\pm 210$ \AA\ and $\lambda_3(0)=3,920\pm 60$ \AA.\\ 

\noindent
{\bf 5. Theoretical Analysis}\\

\noindent
To interpret our data in terms of the pairing symmetry of \UP, we provide a brief discussion of the nodal structures of the superconducting gap. 
Gap profiles in the B-phase for various candidate models for the symmetry of the order parameter are shown 
on an ellipsoidal Fermi surface in the inset to Fig.~\ref{Fig7}b.  In the low field and low temperature B-phase, where all of the data shown in Fig.~\ref{Fig6} were measured, the three predominate pairing models discussed earlier have three different nodal structures.  For the E$_{2u}$ model~\cite{Sauls_AdvPhys_1994}, there are point nodes at the poles of the Fermi surface which open with quadratic wave-vector dispersion,  and a line node around the equator of the Fermi surface that opens with linear dispersion.  The E$_{1g}$ model~\cite{Park_PRB_1996} also has point nodes at the poles, however these nodes open linearly.  Similar to E$_{2u}$, the E$_{1g}$ model also has a line node around the equator that opens linearly.  The E$_{1u}$ model \cite{Tsutsumi_JPSJ_2012} has a somewhat more complicated gap structure in the B-phase, with point nodes at the poles that have 
linear dispersion and two line nodes in planes parallel to the equator where there is an antinode. 
Our measurements of $\lambda_1$ test the nodal structure on parts of the Fermi surface having a significant basal plane component 
of the Fermi velocity, while measurements of $\lambda_3$ are sensitive to the nodes where the  Fermi velocity has a large $c$-axis component.  

For a polar point node with quadratic dispersion, a linear temperature dependence of \lam\ is expected in the low temperature limit, while for a point node with linear dispersion, there would be a $T^2$ temperature dependence.  Extending analysis to a wider range of temperature and interpretation of our results in terms of order parameter symmetry  requires a theoretical calculation including the effects of thermal  excitations of quasiparticles averaged over the whole Fermi surface. \\

We performed calculations of the penetration depth within the framework of the quasiclassical theory~\cite{SR,xu95}, and compared  three models for the symmetry of the order parameter E$_{2u}$, E$_{1g}$, and E$_{1u}$ in Fig.~\ref{Fig7}a,b.
The calculations of superfluid density $\rho_{n}(T)$ and penetration depth $\lambda_{1,3}(T)/\lambda_{1,3}(0)=
\sqrt{\rho_{1,3}(0)/\rho_{1,3}(T)}$ were performed for an ellipsoidal Fermi surface $p_x^2+p_y^2+3p_z^2=p_0^2$, 
with mass anisotropy $m_1/m_3=3$ to account for the observed anisotropy of the penetration depth $\lambda_1(0)^2/\lambda_3(0)^2 \approx 3.3$, 
and the normal state transport, $\kappa_c(T_c)/\kappa_{ab}(T_c) \approx 2.8$~\cite{Lussier_PRL_1994}.
The choice for the form of the order parameter is less obvious, and
there are several approaches to model the gap structure using standard functions, such as spherical functions, or Allen Fermi surface harmonics~~\cite{NormanHirschfeld}. 
We made the more natural choice of ellipsoidal harmonics since they are orthogonal on an ellipsoidal Fermi surface, and 
transform into spherical harmonics with proper rescaling of the Fermi surface. 
Substituting $(p_x, p_y, p_z)=(k_x, k_y, k_z/\sqrt{3})$ we write the gap profiles for the three models,
\begin{eqnarray}
\begin{aligned}
	& \Delta_{E_{2u}} = \Delta_0 | k_z (k_x+i k_y)^2 |\\
	& \Delta_{E_{1g}} = \Delta_0 | k_z (k_x+i k_y) | \\
	& \Delta_{E_{1u}} = \Delta_0 \left| (5 k_z^2-p_0^2) \sqrt{k_x^2 + k_y^2} \right|
\end{aligned}
\label{Eq5}
\end{eqnarray}
displayed in the inset to Fig.~\ref{Fig7}b. 
We used a single-component model for the order parameter, valid deep inside the B-phase and followed the approach from previous work~\cite{NormanHirschfeld}, normalizing the temperature to the lower critical temperature $T_{AB} \approx 0.88 T_c$. 

For this choice of gap functions, and treating $\lambda(0)$ as the \textit{only} adjustable quantity, the theory with $E_{2u}$ order parameter symmetry closely replicates the observed $\lambda_3(T)/\lambda_3(0)$ over a wide range of temperature. The best fit is shown in Fig.~\ref{Fig7}b by a solid blue line, whereas the $E_{1g}$ and $E_{1u}$ models with ellipsoidal harmonics provide considerably worse fits. The $E_{2u}$ model is also consistent with the data for $\lambda_1$ (solid circles) in Fig.~\ref{Fig7}a.

It was pointed out that `pure' ellipsoidal or spherical harmonics might not necessarily 
reflect the correct low-energy structure of the excitations, and do not replicate the observed anisotropy of the heat transport, $\kappa_c(T)/\kappa_b(T)$~\cite{Graf_JLTP_1996}.  A set of 
gap functions was suggested that  are parametrized near line and point nodes with variable slope coefficients $\mu$ for angles $\delta\theta$ with respect to the $c$-axis.
\begin{eqnarray}
\begin{aligned} 
	& \Delta^{line}_{E_{2u}} = \mu  \Delta_0 \delta\theta  \qquad  
		\Delta^{point}_{E_{2u}} = \mu_2 \Delta_0 \delta\theta^2  \\
	& \Delta^{line}_{E_{1g}} = \mu \Delta_0 \delta\theta  \qquad  
		\Delta^{point}_{E_{1g}} = \mu_1 \Delta_0 \delta\theta  
\end{aligned}
\label{Eq6}
\end{eqnarray}
Eq.~\ref{Eq5} corresponds to the opening nodal parameters: $\mu=\mu_1=\mu_2=1$. 
However these authors~\cite{Graf_JLTP_1996} only found a good fit for the thermal conductivity data and sound attenuation with 
$\mu=1$, $\mu_1=1/3$, $\mu_2=2$~\cite{Graf_JLTP_1996,Graf_PRB_2000}.

The latter parameter set results in dashed lines in Fig.~\ref{Fig7}b. With this ansatz the $E_{2u}$ model cannot fit the data over the entire temperature range, while $E_{1g}$ follows the observed data fairly well within error bars but with significantly higher $\chi^2$ value at lower temperatures, as discussed in Fig.~\ref{Fig6}b. This model is also a less likely candidate based on previous analysis of sound attenuation~\cite{Graf_PRB_2000}. In summary, it is compelling that $E_{2u}$ symmetry with a simple parameter set and elliptical harmonics is in excellent agreement with our measured penetration depths over a wide temperature range. \\

\noindent
{\bf 6. Discussion and Summary}\\

\noindent
Signore \textit{et al.}~\cite{Signore_PRB_1995}, reported a linear temperature behavior which could not be associated with any specific component of the penetration depth.  Interpretation of their ac-susceptibility measurements  requires an analysis of the real and imaginary parts of the electromagnetic response from which extraction of the penetration depth is not trivial and is necessarily sensitive to surface quality~\cite{GrossAlltag_ZPB_1991, Signore_PRB_1995}.    An early $\mu$SR investigation by Broholm \textit{et al.}~\cite{Broholm_PRL_1990} found a penetration depth anisotropy much too small to be consistent with  other observations of the superconducting state~\cite{Kleiman_PRL_1992, Yaron_PRL_1997, Joynt_RevModPhys_2002}.  In a later $\mu$SR study, Yaouanc \textit{et al.}~\cite{Yaouanc_JPCM_1998} obtained \lam$_1(0) = 6,040\pm 130$ \AA\  and  \lam$_3(0)= 4,260\pm 150$ \AA\ with $H=$ 0.018 T, qualitatively consistent with what we report here.

Evidence for gap nodes has been sought  from the low temperature behavior of the thermal conductivity and attenuation of sound~\cite{Lussier_PRL_1994, Lussier_PRB_1996, Suderow_JLTP_1997, Ellman_PRB_1996, NormanHirschfeld, Graf_JLTP_1996, Graf_PRB_2000}. %B. Ellman and L. Taillefer, Phys. Rev. B. 54, 9043 (1996).
 The earliest reports~\cite{Lussier_PRL_1994}, provided evidence for both a polar gap node along the $c$-axis and a line node in the basal plane.  However, a conclusion in terms of a specific order parameter symmetry  from nodal gap quasiparticle excitations was not  possible~\cite{Graf_JLTP_1996,NormanHirschfeld}. At high temperatures in the A-phase, measurements of transverse sound attenuation~\cite{Ellman_PRB_1996}, vortex lattice structure~\cite{Huxley_Nature_2000}, and directional tunneling~\cite{Strand_Science_2010} are consistent with  E$_{2u}$ symmetry.   In contrast, a recent report of the directional dependence of the thermal conductivity in the B-phase was argued to support a E$_{1u}$ state~\cite{Tsutsumi_JPSJ_2012}.  This theory requires weak spin-orbit coupling in order to maintain consistency with spin susceptibility measurements from the Knight shift.  The latter is in conflict with observations of Pauli limiting anisotropy evidenced in the upper critical field~\cite{Shivaram_PRL_1986} and it is in conflict with most other theoretical work~\cite{Graf_PRB_2000, Joynt_RevModPhys_2002}.    Our approach has been to use SANS to measure the vortex structure from which we have determined the penetration depth.  These measurements are not compromised by imperfections at the sample surface since they are an average over the whole superconducting crystal and they provide absolute values for the penetration depth.  The interpretation of transport measurements  makes an assumption for the existence of a single order parameter domain that is not required for our measurements of the penetration depth from which we infer that  superconductivity in \UP\ is an odd parity  state with E$_{2u}$ symmetry and  that consequently, the B-phase is chiral.  \\  

\noindent
{ \bf Acknowledgments}\\

\noindent
Research support  was provided by the U.S. Department of Energy, Office of Basic Energy Sciences, Division of Materials Sciences and Engineering under Awards  DE-FG02- 10ER46783 (University of Notre Dame and Northwestern University; neutron scattering) and DE-FG02-05ER46248 (Northwestern University; crystal growth and characterization).  KJS acknowledges support from the Notre Dame Glynn Family Honors program, and JH from the Notre Dame physics REU program. ABV acknowledges support by the National Science Foundation through grant DMR-0954342. This work is based on experiments performed at the Swiss spallation neutron source SINQ, Paul Scherrer Institute, Villigen, Switzerland.  We thank Jim Sauls for his continuing advice and theoretical support throughout the course of the project and  Chris Steiner for his contributions.  We are grateful to PSI for their hospitality and support during this work and we thank M. Zolliker, C. A. Collett, A. Zimmerman, J. I. A. Li and J. Pollanen for their assistance.\\

\noindent
{\bf References}\\

\bibliography{UPt3}

\end{document}